\documentclass[balance,upint,varvw,mathalfa=cal=boondoxo,spanish,french,vietnamese,russian,greek,pdf-a,fontspec,colorlinks,nofoot]{asmeconf}

\hypersetup{%
	pdfauthor={Florian Stadtmann, Henrik Gusdak Wassertheurer, Adil Rasheed},
	pdftitle={Demonstration of a Standalone, Descriptive, and Predictive Digital Twin of a Floating Offshore Wind Turbine},                  
	pdfkeywords={Digital Twin, Offshore Wind, Floating Offshore Wind Turbine},
	pdfsubject = {Demonstration of a Standalone, Descriptive, and Predictive Digital Twin of a Floating Offshore Wind Turbine},			  
}

\usepackage{url}
\urldef{\urla}\url{https://www.youtube.com/watch?v=FswHdSx_mOQ&t=8s}
\urldef{\urlb}\url{https://openphotographyforums.com/forums/threads/wind-power-in-west-texas.14738/}

\usepackage[flushmargin,ragged]{footmisc}

\begin{document}



\title{Demonstration of a Standalone, Descriptive, and Predictive Digital Twin of a Floating Offshore Wind Turbine} 
 
%
%
%

\SetAuthors{%
	Florian\ Stadtmann\affil{1}\CorrespondingAuthor{florian.stadtmann@ntnu.no}, 
	Henrik Gusdal Wassertheurer \affil{1}, 
	Adil Rasheed\affil{12} 
	}

\SetAffiliation{1}{Norwegian University of Science and Technology NTNU, Trondheim, Norway}
\SetAffiliation{2}{Department of Mathematics and Cybernetics, SINTEF Digital, Trondheim, Norway}


\maketitle





\begin{abstract}
Digital Twins bring several benefits for planning, operation, and maintenance of remote offshore assets. 
In this work, we explain the digital twin concept and the capability level scale in the context of wind energy. Furthermore, we demonstrate a standalone digital twin, a descriptive digital twin, and a prescriptive digital twin of an operational floating offshore wind turbine. The standalone digital twin consists of the virtual representation of the wind turbine and its operating environment. While at this level the digital twin does not evolve with the physical turbine, it can be used during the planning-, design-, and construction phases. At the next level, the descriptive digital twin is built upon the standalone digital twin by enhancing the latter with real data from the turbine. All the data is visualized in virtual reality for informed decision-making. Besides being used for data bundling and visualization, the descriptive digital twin forms the basis for diagnostic, predictive, prescriptive, and autonomous tools. A predictive digital twin is created through the use of weather forecasts, neural networks, and transfer learning. Finally, digital twin technology is discussed in a much wider context of ocean engineering.

\end{abstract}


\begin{nomenclature}[3em]
\entry{AR  }{Augmented reality}
\entry{CAD }{Computer-aided design}
\entry{DNN }{Dense neural network}
\entry{DOF }{Degrees of freedom}
\entry{DT  }{Digital twin}
\entry{GUI }{Graphical user interface}
\entry{HAM }{Hybrid analysis and modeling}
\entry{LSTM}{Long-short-term-memory neural network}
\entry{NN  }{Neural network}
\entry{O\&M}{Operation and maintenance}
\entry{RPM }{Revolutions per minute}
\entry{VR  }{Virtual reality}
\end{nomenclature}


\section{Introduction}
    \label{sec:intro}
    Offshore wind energy production will significantly increase in the coming years~\cite{2018ida,Freeman2019oeo}. Especially floating offshore wind farms allow for tapping the enormous wind resources available in deep waters~\cite{IEA2019owo}. However, their remote locations result in several challenges connected to planning, operation, and maintenance, including a harsh environment, a potential lack of local expertise, expensive on-site inspections, and long response time to unexpected downtime \cite{Adumene2022oss}.
    
    The digital twin (DT) concept is an upcoming technology that addresses these challenges by making it possible to perform significant parts of the operation and maintenance (O\&M) process remotely. However, the understanding of what features precisely are and are not part of a DT varies \cite{Valk2020ATO},
    and conventional scales, such as the technology readiness levels, cannot account for these varying capabilities of the DT.
    Recently, a capability level system has been proposed that allows ranking any DT on a scale from 0 to 5 (0-standalone, 1-descriptive, 2-diagnostic, 3-predictive, 4-prescriptive, and 5-autonomous)~\cite{DNVGL2020dra,Sundby2021gcd}.
    However, the implementation of DTs has not been sufficiently explored for offshore applications on any of these capability levels, and their meaning and value have not been specified for offshore assets and wind energy in particular.
    Furthermore, through communication with the Norwegian wind energy industry, it has been discovered that there is a lack of documentation on DTs and of industrially relevant use cases.
    
    To this end, the article aims at:
    \begin{itemize}
        \item familiarizing the reader with the DT concept and the recently proposed capability level scale.
        \item explaining the value that DTs bring for remote offshore assets and wind turbines in particular.
        \item demonstrating the standalone, descriptive, and predictive capability of a DT as a use case for an existing offshore floating wind turbine.
        \item giving an outlook for future developments on DTs for remote assets.
    \end{itemize}
    
    Section~\ref{sec:definition} gives the definition of the DT concept for the purpose of this work and familiarizes the reader with the capability level structure.
    In Sec.~\ref{sec:method} the components required for a standalone, descriptive, and predictive DT are elucidated on a general level.
    In Sec.~\ref{sec:implementation} their implementation is explained.
    Section~\ref{sec:discussion} summarises this work, discusses further challenges that have to be addressed to enable DTs on all capability levels, and provides an outlook toward future work. Additionally, the expected impact of DTs on ocean engineering is discussed.
    In Sec.~\ref{sec:conclusion}, a short conclusion of the paper is given.
    
\section{Digital Twins and Capability Level}
    \label{sec:definition}
    There has previously been a lack of consensus within the industry and research as to what a DT is, and many attempts have been made to distill a singular definition, e.g. in~\cite{Solman2022dta}.
    To avoid any misconceptions, the term \textit{digital twin} (or DT) in this work is defined as a virtual representation of a physical asset enabled through data, models, and analysis for real-time prediction, optimization, monitoring, control, and informed decision-making (compare~\cite{Rasheed2020dtv}).
    The DT concept can be used for a variety of tasks, and not all tasks require all the features mentioned above. To determine what features are and are not implemented in a DT, this work adapts the capability level system from~\cite{DNVGL2020dra,Sundby2021gcd}, specifies it further, and explains its meaning for offshore assets and wind energy applications.
    Therefore, a DT can be \textit{"ranked on a scale from 0 to 5 (0-standalone, 1-descriptive, 2-diagnostic, 3-predictive, 4-prescriptive, and 5-autonomous)."}. The different capability levels will be elaborated on in the following, and their application in the context of wind turbines and other offshore assets will be explained.
    
    \begin{figure}[h]
        \centering
        \includegraphics[width=\linewidth]{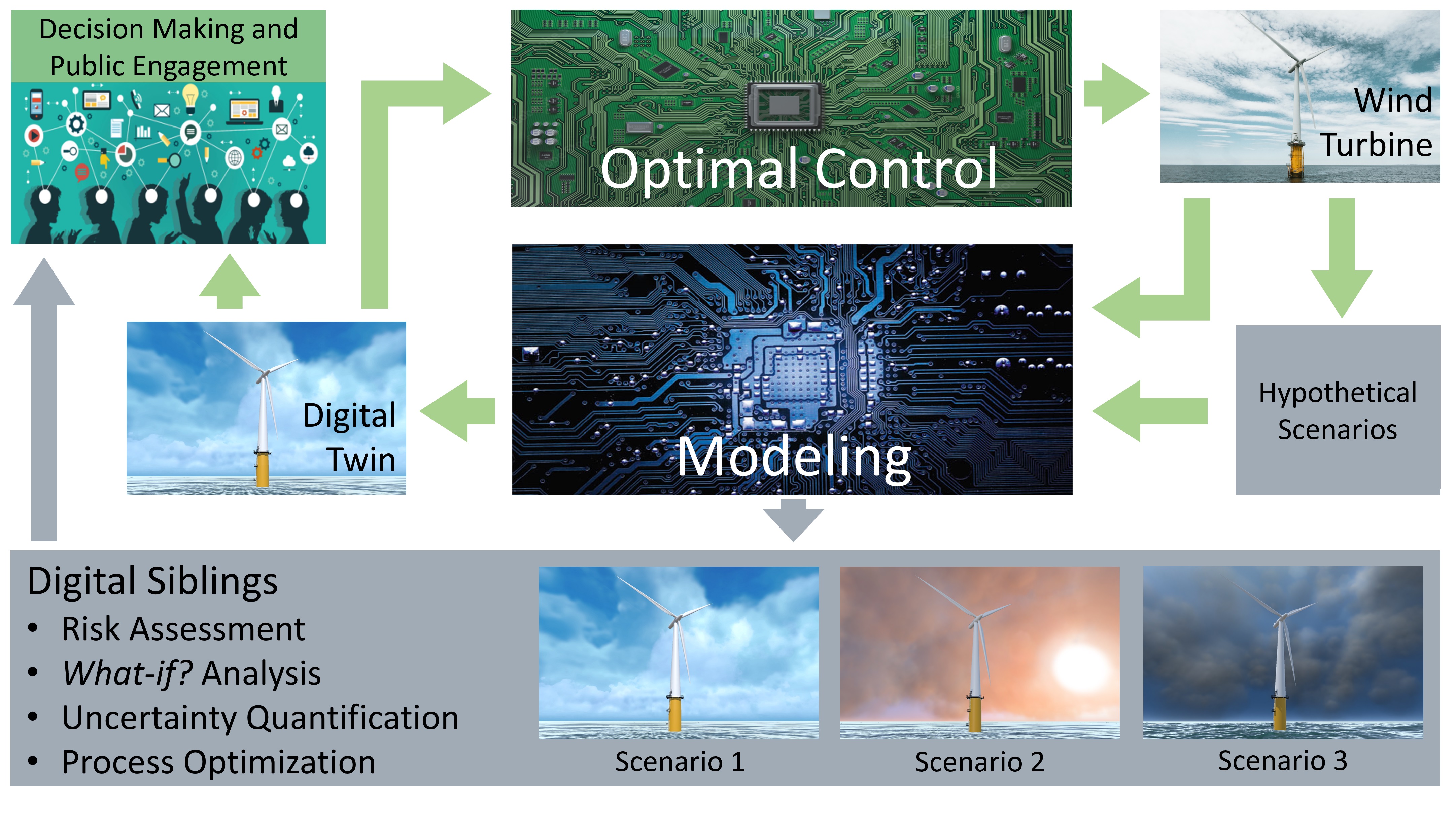}
        \caption{Digital twin framework for the test site, adapted from~\cite{Pawar2021haa}}
        \label{fig:dt}
    \end{figure}
    
    \begin{figure*}[h]
        \centering
        \includegraphics[width=\linewidth]{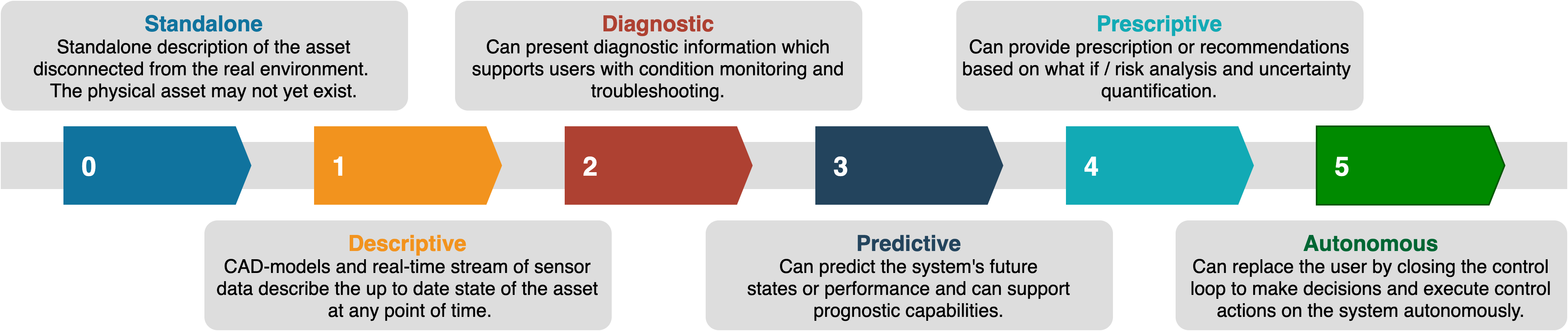}
        \caption{Capability level of any digital twin, adopted from~\cite{Sundby2021gcd}}
        \label{fig:capability}
    \end{figure*}
    
    \subsection*{Standalone Digital Twin}
        \label{ssec:d0}
        A standalone DT is a representation of a physical asset that has no real-time data connection to the asset. Generally, the physical asset does not need to exist for this to be implemented. This can constitute a 3D model of the asset and its environment. 
        In the context of a floating offshore wind turbine, the relevant environment could be the ocean and ocean floor. The standalone DT can be used for planning tasks such as wind turbine siting or impact factor assessment. Furthermore, the model can be used for the design of turbine components, for research and development, and for displaying a comparatively real-size illustration to stakeholders or potential investors.
    
    \subsection*{Descriptive Digital Twin}
        \label{ssec:d1}
        A descriptive DT adds real-time updates through sensors to the model and thus describes the physical asset at any given point in time. This allows for an accurate representation of real-life conditions in the DT and can be used to monitor the physical asset remotely. 
        In the case of floating offshore wind turbines, the visualization data can constitute everything from e.g. the turbine rotation, pitch, yaw, tower movement, temperature levels, pressure levels, vibration, power production, wave spectrum, atmospheric data, availability, and functional capacity of individual parts. The descriptive DT allows for the packaging of a large amount of data into an easily observable form and can provide simplified explanations for complex interactions between the different data that are easily spotted when visualized. The visualization also provides a data bundling and visualization platform that can be used for further analysis at higher capability levels with both historic and real-time data.
    
    \subsection*{Diagnostic Digital Twin} 
        \label{ssec:d2}
        The DT gains diagnostic capabilities when the DT is enhanced with analysis tools that notify the user about unusual asset behavior and help with troubleshooting. For example, condition monitoring features can be implemented that keep track of e.g. vibration of components, raise warnings when critical thresholds are surpassed, and indicate the damage to the component inside the DT. It is essential in realizing condition-based maintenance of any asset, which is especially important for remote assets like floating offshore wind turbines where in-person inspections are expensive to perform.
    
    \subsection*{Predictive Digital Twin}
        \label{ssec:d3}
        The predictive DT aims at estimating the future state of the system to support operation, maintenance, and more. 
        The prediction of future states can help schedule appropriate downtime for testing or maintenance to optimize the operational time of the asset. It can decrease the likelihood of excessive downtime due to critical failure. For wind farms, it can give humans that are in the loop the opportunity to balance the power output against component wear in real-time and get optimal power production predictions for the day-ahead energy market.
    
    \subsection*{Prescriptive Digital Twin}
        \label{ssec:d4}
        A prescriptive DT provides recommendations based on what-if scenarios and risk analysis and thereby aids with decision support. In the context of wind turbines, this can help optimize the power output, schedule maintenance, and avoid excessive maintenance and downtime by identifying optimal maintenance strategies. It can also show potential risks of economic loss from downtime and from the damage caused by not delaying maintenance after faults have been identified. This allows the human controller to assess which decisions are economical and sustainable.
    
    \subsection*{Autonomous Digital Twin}
        \label{ssec:d5}
        The autonomous DT replaces the human in the loop. The autonomous capability can vary from an enhanced control system in the DT to a partially or fully autonomous system with respect to operation and possibly maintenance. For the latter case, drones can inspect the turbine blades, the interior, and the underwater structure. At later stages, autonomous vehicles could be included to perform the necessary maintenance operations and transport spare parts, to the point where eventually the asset operates and maintains itself.

\section{Components of the Digital Twin}
    \label{sec:method}
    In this section, the essential components are explained that enter into the creation of a DT at each of the three capability levels demonstrated in this work. First, a general explanation is given here, and afterward, the details of their implementation are presented Sec.~\ref{sec:implementation}.
    \subsection{Standalone Digital Twin}
        \label{ssec:m0}
        The standalone DT requires a suitable visual representation of the asset (likely a 3D model). Furthermore, it requires software that can render the model in its environment, enables interaction with the model, and allows the implementation of physically correct behavior.
        \subsubsection*{3D Model}
            \label{sssec:m0 3d}
            A 3D model of the physical asset serves as the basis for the standalone DT and provides an advanced visualization platform for the human-machine interface of higher capability levels. Often, computer-aided design (CAD) models of single components are created during the design of said components. These component models can be combined to create the full asset model. However, in cases where no models are available or proprietary conflicts arise, new CAD models need to be created. It is important to note that the CAD models must not take up too much of the available computational resources when rendered. A trade-off between the number of vertices/triangular surfaces in the model and the technical detail of the model must be made. Therefore the way vertices are being used in the model needs to be optimized. Already existing CAD models can still be used by applying automated vertex reduction.
        \subsubsection*{Game Engine}
        \label{sssec:m0 engine}
            The CAD model needs to be imported into a 3D computer graphics engine that allows rendering, modifying, and exploring the model, implementing relevant physics, creating visualization tools, and building an interface. There are several engines that are capable of running detailed physical simulations. However, especially for higher capability levels, it is crucial that the DT can be executed in real-time. Additionally, it is desirable to choose and implement physics for each component according to the required precision and replace some physics-based equations with data-driven or hybrid analysis and modeling (HAM) approaches to improve performance and accuracy.
            Such flexible engines have been used in the computer game industry for decades and have also been adopted for engineering tasks~\cite{Gang2016aso, Li2022vms}.
            
        \subsubsection*{Environment}
        \label{sssec:m0 environment}
            In many situations, interaction with the environment is highly important. For offshore assets, the environment may comprise wind and weather, waves, ocean depth, sea bed, coastline, or even other offshore assets. The environment can be built in the engine and populated with relevant data. As with the CAD model of the asset, it is important that the computational efficiency is not compromised by the implementation of the environment. 

        \subsubsection*{Virtual and Augmented Reality}
            \label{sssec:m0 VR}
            While it is sufficient to have the DT set up for usage through a regular computer screen, it can be beneficial to use virtual reality (VR) or augmented reality (AR) for a more intuitive understanding of - and improved interaction with - the DT.
            VR allows the user to explore the DT as if interacting with the real asset. In the standalone DT, it is possible to interact with the asset and its components even before they are built, not only providing a great tool for engineers but also easing communication with non-technical stakeholders who may not be able to interpret the technical 2D sketches conventionally used. The VR implementation stays a valuable tool also in higher capability levels, e.g. to train technical personnel and plan maintenance in the DT, which can improve security, speed, and precision of maintenance on site. Furthermore, the combination of model and data allows remote inspections of offshore or arctic assets even during harsh weather conditions without leaving the office.
            While VR has strong applications for remote operations, AR can be beneficial for on-site operations. By overlaying the real asset with the DT, information can be provided next to the real components and analytically uncovered flaws invisible to the human eye can be visualized to the maintenance crew.
    
    \subsection{Descriptive Digital Twin}
        The descriptive capabilities of a DT are unlocked when it is populated with real-time data. A data pipeline is needed to manage the real-time data flow between the asset, the computational unit (server or cloud), the data storage, and all clients/user-interfaces.
        \label{ssec:m1}
        \subsubsection*{Data Pipeline}
        \label{sssec:m1 pipeline}
\begin{figure}
            \centering
            \includegraphics[width=\linewidth]{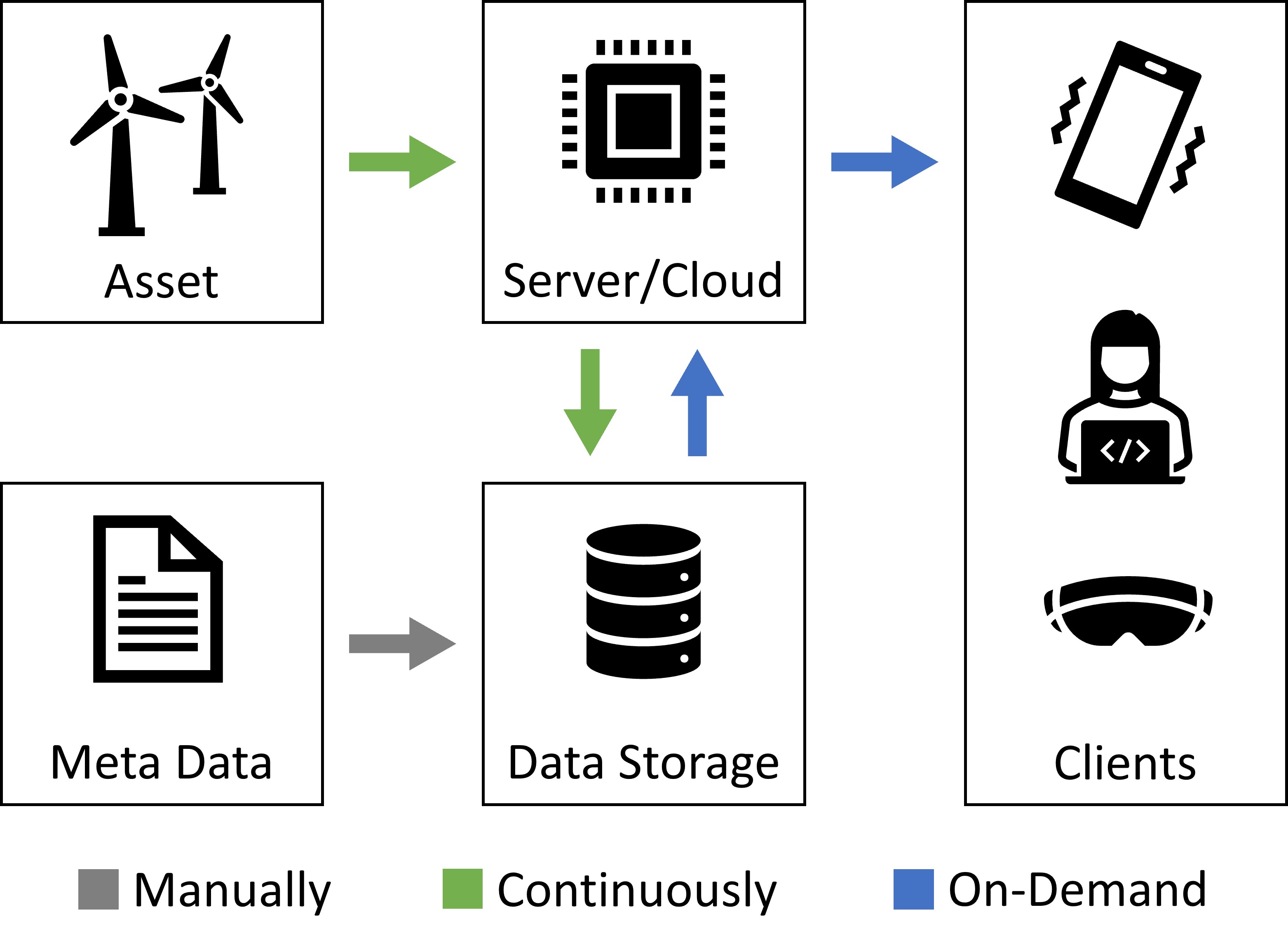}
            \caption{Data pipeline of a descriptive to prescriptive digital twin. At the autonomous level, feedback from the client and server/cloud to the asset will be added.}
            \label{fig:pipeline}
        \end{figure}

            A simplified DT data pipeline is shown in Fig.~\ref{fig:pipeline}. Green arrows symbolize real-time data transfer, while blue arrows show on-demand data transfer (e.g. through HTTPS requests). Manually implemented data such as metadata is shown in grey.
            The data measured at the wind turbine/farm is sent to shore in real-time. There, a server or cloud-based computing system receives the data. In the descriptive DT, the server decides what data to send to the data storage, and compresses the data where it has not been done at the asset site already. On higher capability levels, the data is analyzed at the server immediately for the purpose of condition-based maintenance, predictive maintenance, what-if scenario analysis, and - at the autonomous stage - also control. Anomalies are detected and alerts are sent to the clients automatically.
            At any time, the clients can request real-time data and historic data from the server through HTTPS requests.
            The client can be a desktop computer, a laptop, a mobile phone, or a VR/AR headset. 
            Simple on-demand analysis and predictions can be performed on the user's device, while the intermediate results of more complex analysis are requested from the server/cloud. 

    \subsection{Predictive Digital Twin}
        \label{ssec:m3}
        In the predictive DT, the future state of the asset and its environment is predicted. In some cases, forecasts are already available from external sources, but asset-specific parameters have to be predicted internally by the DT. To this end, physics-based and data driven-methods can be used separately or can be combined into HAM approaches.
        \subsubsection*{External}
            \label{ssec:m3 prediction}
            Especially predictions related to the environment, e.g. weather forecasts, are available from external sources. Those forecasts can be directly utilized in the DT to predict the weather but also to infer asset behavior. For wind turbines, weather forecasts can be used to calculate forecasts of power production.
        \subsubsection*{Physics-based}
            Physics-based models can provide accurate and reliable predictions of specific quantities by using e.g. differential equations for wind fields~\cite{Rasheed2014amw}. 
            However, in some cases, the physics-based model is not fast enough to execute. Reduced order models are an approach that aims at reducing computational cost and increasing efficiency of the physics base predictions, but sacrifice accuracy for it. Furthermore, domain knowledge is needed for every parameter. In complex DTs with dozens or hundreds of different parameters, maintaining the speed and interoperability of the full system can become very challenging.
        \subsubsection*{Data-driven}
            Data-driven methods provide the opportunity to significantly increase the evaluation speed and allow adaption to asset-specific intricacies. Machine learning approaches, especially neural networks (NNs) like dense NNs (DNN), convolutional NNs, or recurrent NNs, have been used not only for classification and regression but also for prediction models. Especially long-short-term-memory neural networks (LSTM), a specific form of recurrent NNs, have been used for forecasting series due to their ability to infer both short- and long-term dependencies through memory cells. However, while NNs excel at speed, they need large amounts of data to train and they can be unreliable, as evident for example from adversarial attacks~\cite{Long2022aso}.
        \subsubsection*{Hybrid Analysis and Modeling}
            HAM summarizes approaches that combine physics-based and data-driven methods.
            They aim at combining the speed and adaptability of data-driven methods with the reliability of physics-based models e.g. by combining NNs with reduced order models.

\section{Implementation}
    \label{sec:implementation}
    In this section, it is explained how the standalone, descriptive, and predictive DTs were implemented for the floating offshore wind farm Zefyros specifically. Nonetheless, the procedure can be easily generalized to many other offshore assets. 
    Note that explaining the countless code-specific implementations would exceed the scope of this article by far. To this end, only the implemented modules are explained here, along with the external tools used.
    \subsection{Standalone Digital Twin}
        \label{ssec:i0}
        \subsubsection*{CAD Modeling and Texturing}
            \label{ssec:i0 model}
\begin{figure}[h]
                \centering
                \includegraphics[width=\linewidth]{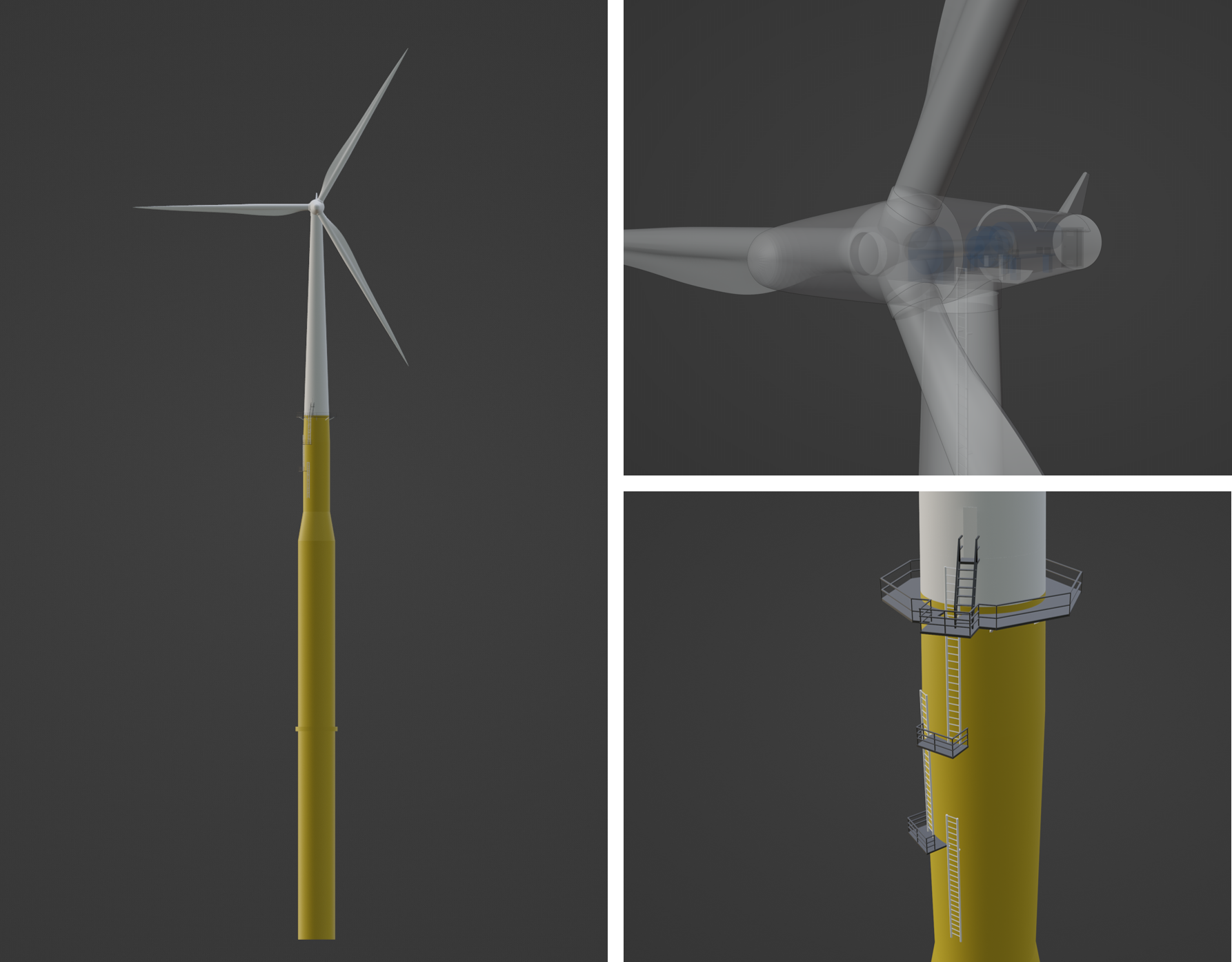}
                \caption{A CAD model of the floating wind turbine. Left: Full Model. Top right: Nacelle. Bottom right: Platform.}
                \label{fig:cad}
            \end{figure}

            No CAD model was available from the floating offshore wind turbine. Therefore, a model had to be created using only openly available images and sketches of the turbine. For the floating structure and tower section heights and diameters, the technical sketch of Zefyros (former Hywind Demo) from~\cite{Keiser2014bp1} was used.
            The platform was designed by combining several images with a video\footnote{\urla}.
            The nacelle geometry and parameters were estimated from pictures\footnote{\urlb}\footnote{\url{https://www.uib.no/sites/w3.uib.no/files/attachments/hywind_energy_lab.pdf}}.
            The turbine itself was identified as the Siemens SWT-2.3MW-82 turbine. Pictures and sketches\footnote{\url{https://en.wind-turbine-models.com/turbines/341-siemens-swt-2.3-101}} were used to create a simplified model of the rotor hub, drive shaft, gearbox, and generator. As no reference airfoils of the turbine blades are openly available, the CAD model of the blades is inspired by pictures only. The texturing of each of the components is done with materials created from scratch based on the images mentioned above. The CAD model was designed in the Blender\footnote{\url{https://www.blender.org/}} software using a minimal vertices approach. The final CAD model is shown in Fig.~\ref{fig:cad}.

        \subsubsection*{Engine}
            \label{ssec:i0 engine}
            In this project, the Unity\footnote{\url{https://unity.com/}} has been chosen as the game engine to host the interface for its wide usage, rich documentation, and good support. The engine allows using the programming language C\# to manage data, add physics, and implement dependencies between model components.
        \subsubsection*{Environment}
            \label{ssec:i0 environment}
            The 3D model was placed in an ocean. For its visualization, the free version of the Ocean Crest Unity asset\footnote{\url{https://github.com/wave-harmonic/crest}} was used. 
            For the sky, a combination of the Oculus Interaction SDK asset \footnote{\url{https://developer.oculus.com/documentation/unity/unity-isdk-interaction-sdk-overview/}} skybox and the SkyShphere Volume 1 asset \footnote{\url{https://assetstore.unity.com/packages/2d/textures-materials/sky/skysphere-volume-1-4042}} is being used.
        \subsubsection*{Interface}
            \label{ssec:i0 interface}
            While the whole engine can be seen as a human-machine interface, a dedicated graphical user interface (GUI) is implemented that looks like a conventional user interface but behaves as a 3D canvas for improved VR integration. Using the GUI, the user can modify the settings and toggle features of the DT. 
        \subsubsection*{Virtual Reality}
            \label{ssec:i0 VR}
\begin{figure}
            \centering
            \includegraphics[width=\linewidth]{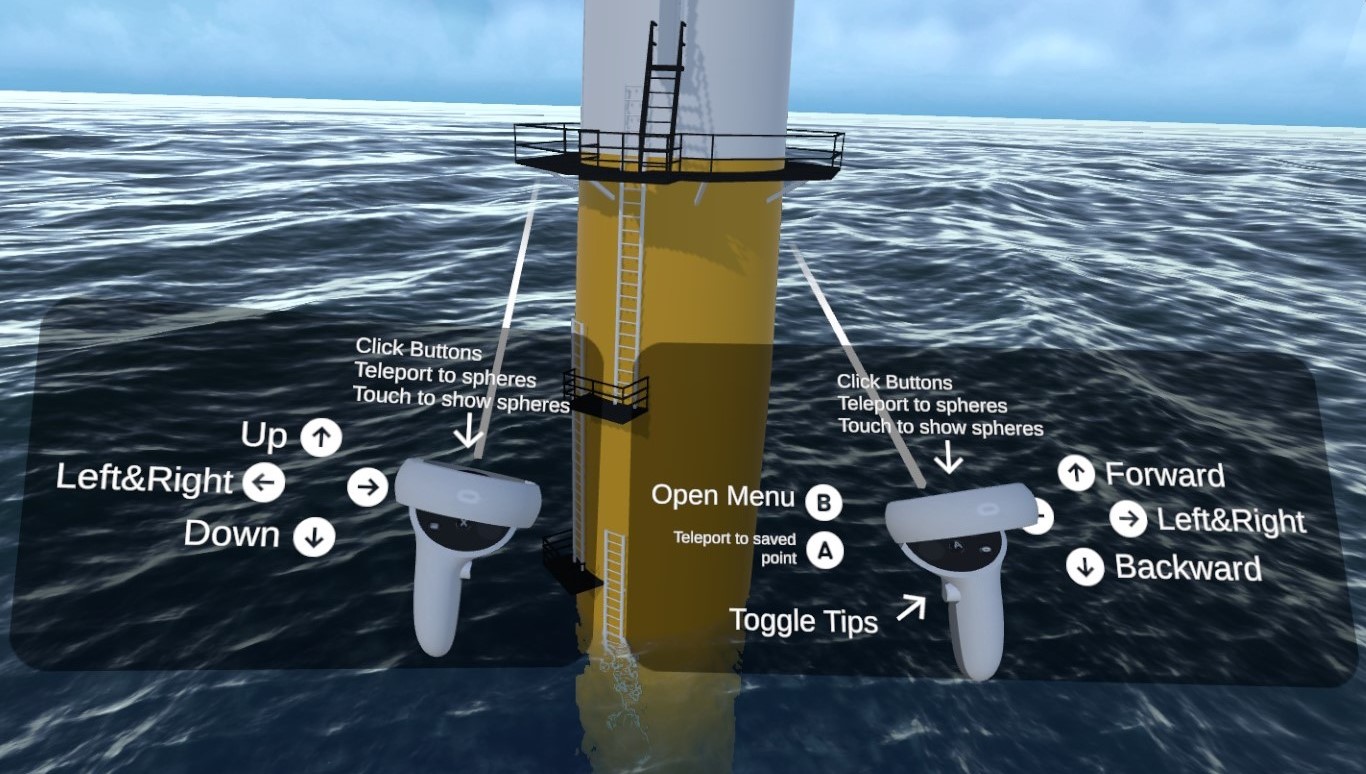}
            \caption{VR controllers with tooltips.}
            \label{fig:controls}
            \end{figure}

            For this project, a VR implementation is being chosen. The VR headset used is the Meta Quest 2. It comes with two controllers, which can be used to interact with the DT. The controls implemented are shown in Fig.~\ref{fig:controls}. Some features can also be accessed without the controller through gesture recognition, but this component is still under development. The integration into Unity happens through the Oculus Interaction SDK asset.
            There are four ways of movement implemented. First, the user can teleport to pre-determined positions inside and outside the turbine by pressing a button to quickly move around. Second, the user can move around using the joysticks on the controller. Since the second option can potentially cause motion sickness for some people, the third option allows teleportation to spots by pointing at them. In combination with the previous three methods, the user's movement in the real world is copied into the DT, allowing the user to walk around in the real world and VR environment simultaneously. Meanwhile, the standard VR guardian system ensures that the user does not walk into real-world objects.
            The interface can be opened with a button and navigated by pointing at its content. Furthermore, a button allows toggling user tips for operation.

        \subsubsection*{Setup Structure}
            \label{ssec:i0 structure}
            The standalone setup consists of several components and scripts. Showing a complete hierarchy would go far beyond the scope of this article, but a very brief summary of the most important components is provided here. In general, the standalone DT implementation can be split into these components:
            \begin{itemize}
                \item The \textbf{Inspector} contains the camera that renders the user's vision and the VR controllers. Additionally, it contains all scripts responsible for movement, the teleport points, and the interactive instruction manual.
                \item The \textbf{Wind Farm} includes the wind turbines (in this implementation only one turbine) with all the CAD models of their components, their relations, positions, and all scripts that make the turbine and its components interactable. This includes e.g. blade pitch, yaw, and rotor rounds per minute (RPM) which can be adjusted in the standalone DT, but also simple scripts for opening doors/hatches.
                \item The \textbf{Environment} comprises the sea floor, the sun/light, and the sky, as well as the ocean surface and all scripts necessary to affect its parameters such as wave height and direction.
                \item The \textbf{GUI} contains its full hierarchy, as well as all executables that the GUI's components trigger directly such as modification of settings and toggling of functionalities. Additionally, it includes the scripts for interaction between GUI and VR controller.
            \end{itemize}

    \subsection{Descriptive Digital Twin}
        \label{ssec:i1}
        \subsubsection*{Available Data} 
            \label{ssec:i1 available}
            The data used in this work was measured from 2.2022 to 9.2022 at the floating offshore wind turbine Zefyros. The measurement frequency depends on the sensor type and varies between 1 to 4 seconds intervals. In total, 58 different parameters and status messages are represented in the data set. The parameters are grouped into logical nodes according to the standard IEC 61850, with the logical node names being an extended version of the information model for wind power plants as in IEC 61400-25. The logical node names and parameters they contain in this work are:
            \begin{itemize}
                \item \textbf{WMET} 7 met-ocean parameters for wind, waves, and temperature
                \item \textbf{WROT} 4 parameters for blade pitch and rotor RPM
                \item \textbf{WYAW} 2 parameters for yaw angle and system status
                \item \textbf{WTOW} 6 parameters for 6 degrees of freedom (DOF) tower movement
                \item \textbf{WTRM} 5 parameters related to temperature and status of the shaft bearing, brakes, and gearbox oil
                \item \textbf{WTUR} 7 parameters for active and reactive power, generator and stator temperature, and turbine operation and status codes
                \item \textbf{WGEN} 2 parameters for generator status and RPM
                \item \textbf{WCNV} Generator frequency
                \item \textbf{WTRF} 10 parameters for transformer 3-phase current, phase-to-phase voltage, as well as transformer oil status and temperature
                \item \textbf{WSTR} Ballast depth
                \item \textbf{WPPD} Control system status
                \item \textbf{WAVL} 12 operation-related parameters, including availability time, operation time, accumulated energy, grid fault time, and several other status messages
            \end{itemize}
            
        \subsubsection*{Data Pipeline and Pre-processing}
            \label{ssec:i1 pipeline}
            Ideally, the full pipeline explained in Sec.~\ref{sssec:m1 pipeline} and shown in Fig.~\ref{fig:pipeline} would be used. However, the data-sharing agreement for this work ended in September 2022. Therefore, no real-time streaming from the turbine can be established. To demonstrate the streaming process, some of the data is being streamed in real-time through HTTPS requests as mentioned in Sec.~\ref{ssec:i3}. The process is extendable to stream from any HTTPS server or through cloud service APIs.
            
            The raw data was provided in a table unsorted with respect to time. To be able to treat the data as if measured in real-time, the data was sorted by time.
            Following this, the data is handled like real-time data. Outlier removal was only performed for unphysical values. For network training and visualization purposes, the data is resampled to 1-second intervals. Linear and cubic interpolations are not an option for real-time usage, as they would use "future" values. Therefore, forward padding was used in this work to infer missing values. Kalman filters are an alternative to padding and allow for smoother visualization, but their implementation requires parameter-specific models and is left for future work.

            \begin{figure}[h]
                \centering
                \includegraphics[width=\linewidth]{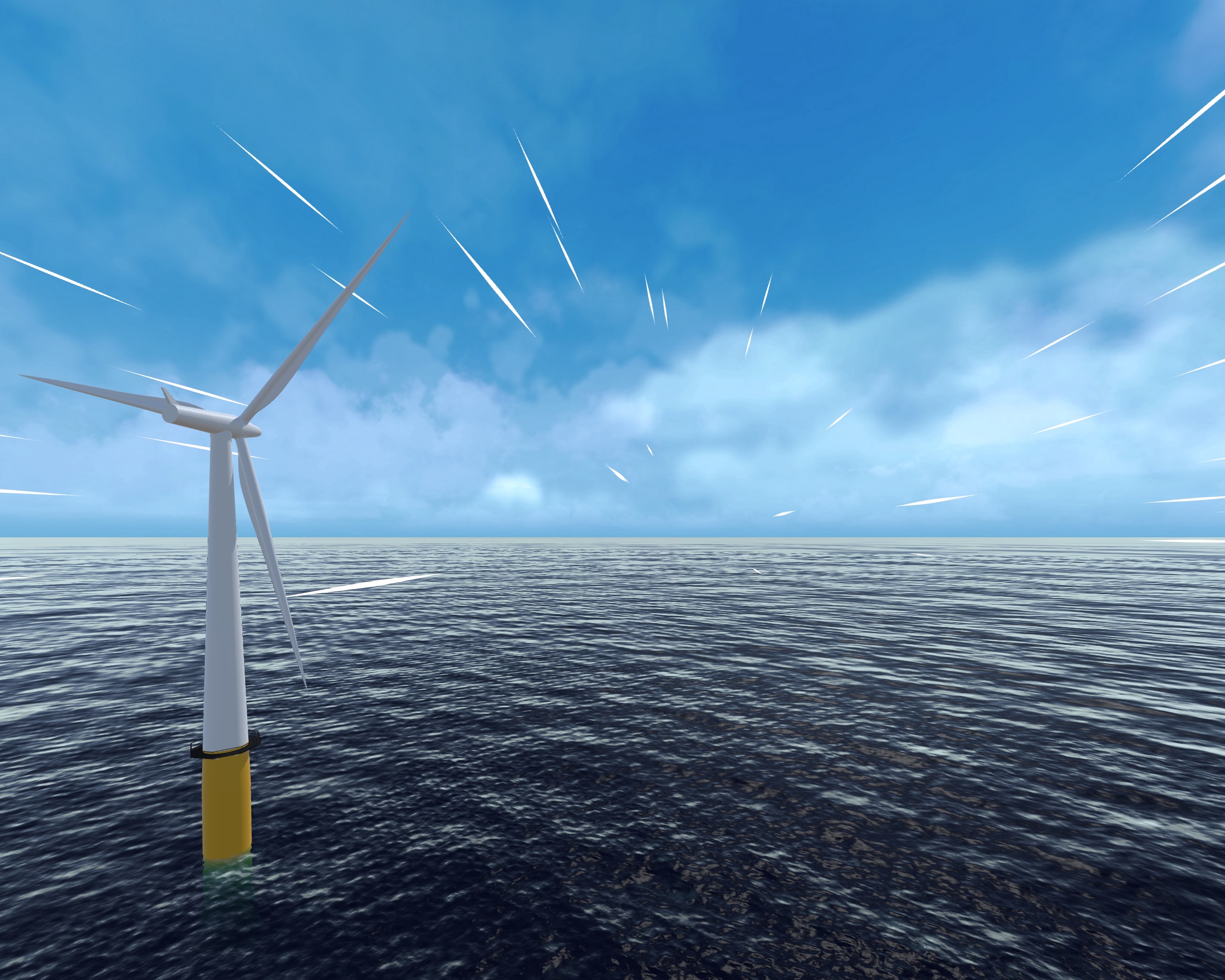}
                \caption{Predicted wind speed and direction are indicated by wind trails moving through a vector field.}
                \label{fig:trails}
            \end{figure}
            
        \subsubsection*{Data Visualization}
            \label{ssec:i1 visualization}
\begin{figure}[h]
                \centering
                \includegraphics[width=\linewidth]{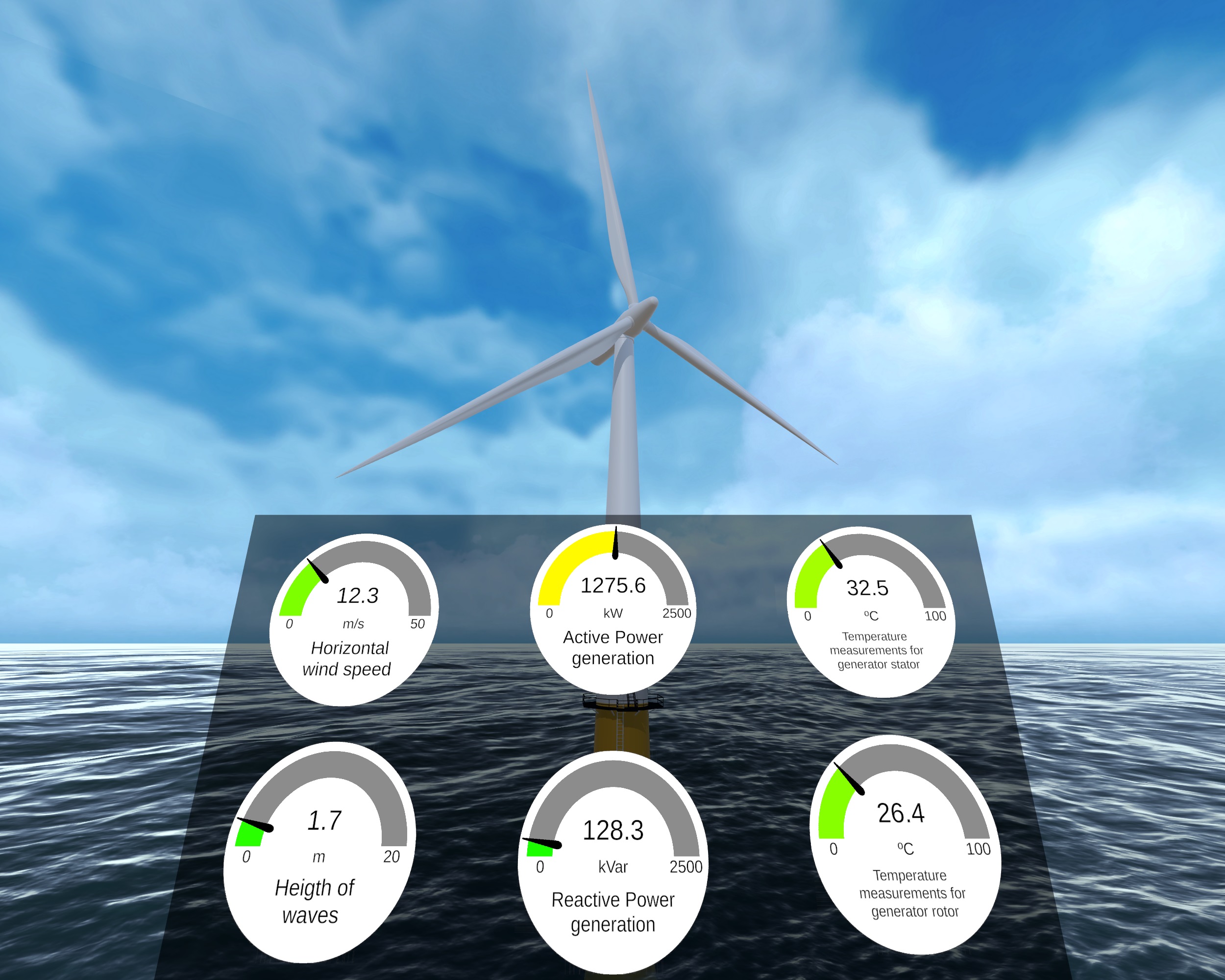}
                \caption{Gauges show important operational conditions.}
                \label{fig:gauges}
            \end{figure}

\begin{figure}[h]
                \centering
                \includegraphics[width=\linewidth]{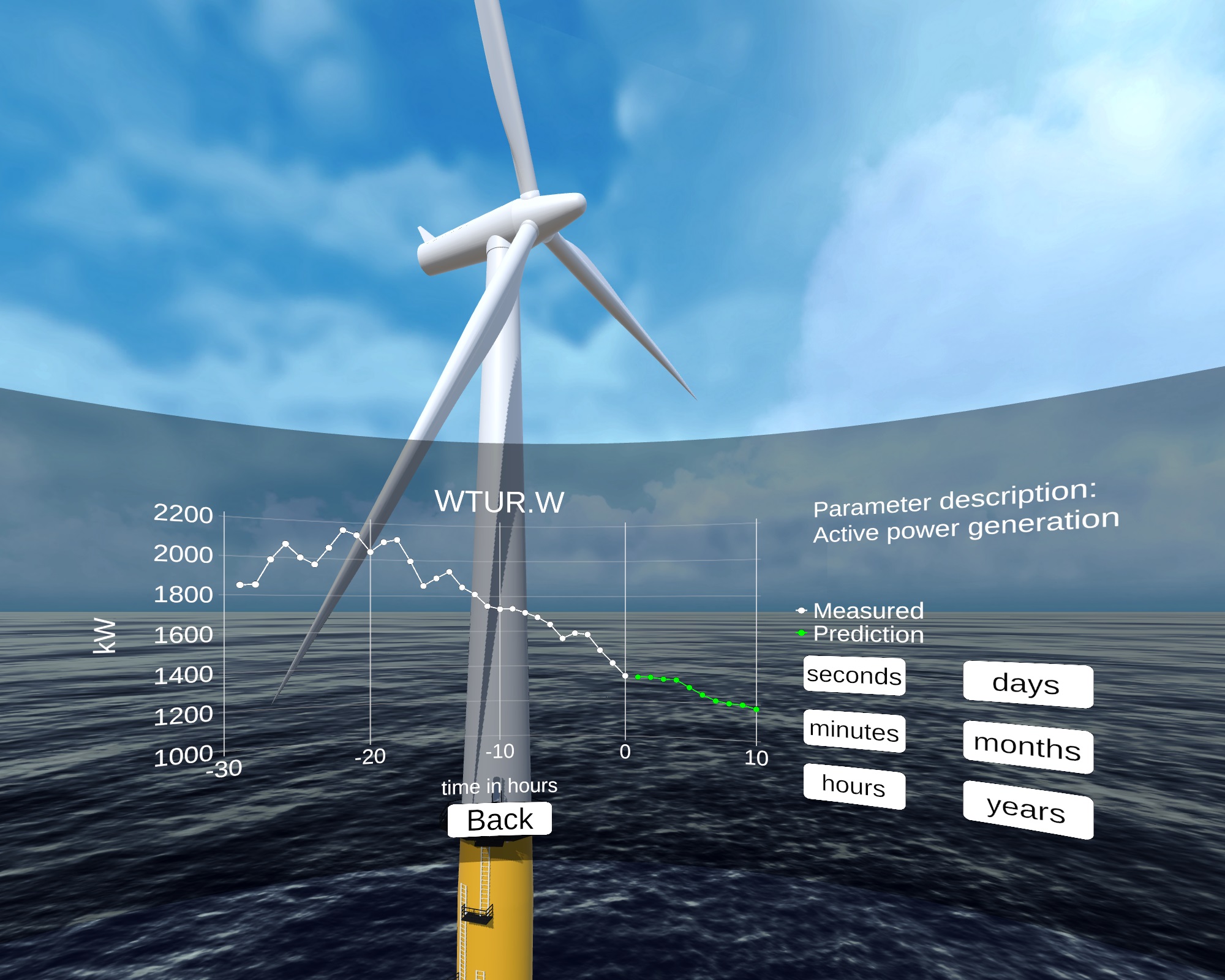}
                \caption{Historic, real-time, and predicted data plotted graphically.}
                \label{fig:graphs}
            \end{figure}

            The visualization of all historical and real-time data happens inside the Unity application. Many parameters can be visualized directly on the turbine by modifying positions and rotations. The 3D model moves according to the measured 6DOF at the tower. Yaw angle and blade pitch are updated with the measured angles. The rotor and generator rotations correspond to the measured RPM, and the ocean wave height is inferred from the data. Furthermore, the measured and forecasted wind can be shown through a trail system, where wind trails move through the vector field and follow the local direction and speed of the wind, as shown in Fig.~\ref{fig:trails}. For a quick overview of operational parameters, a control panel with gauges can be used as shown in Fig~\ref{fig:gauges}. Figure~\ref{fig:graphs} shows an interface where historic, real-time, and forecasted data can be plotted.
            Note that these are only examples of how data can be visualized in a descriptive DT. More approaches are discussed in Sec.~\ref{sec:discussion} and will be explored in future work.

        \subsubsection*{Setup Structure}
            \label{ssec:i1 structure}
            Inside the Unity system, two new components are introduced in addition to the components of the standalone DT to allow for descriptive capabilities:
            \begin{itemize}
                \item The new \textbf{Time} component keeps track of all relevant time parameters, including the real-time, the pretended real-time for demonstrating the DT system on historical data (system time, smaller or equal to real-time), and the time that the user wants to explore the DT in (simulation time, can be the past, present or future). Furthermore, the time component manages the simulation speed for the purpose of time-lapse or slow-motion. A separate animation speed allows all moving parts based on velocities to be independent of time-lapse. Finally, it contains several internal time-related parameters that have to be accessed by many other scripts simultaneously.
                \item The \textbf{Data Loader} component contains the scripts responsible for loading the measured data, as well as streaming the most recent weather data. 
                \item A \textbf{Control Panel} contains the geometries for the gauges and the scripts that read the data and modify the dials accordingly. 
            \end{itemize}
            Furthermore, several upgrades are being made to the standalone structure for the visualization of the data:
            \begin{itemize}
                \item All \textbf{Wind Farm} scripts are modified to read their parameters from a turbine controller, which contains the most recent measurements, interpolation, or extrapolation based on settings. Additionally, scripts are added that show relevant data on the turbine in text.
                \item The \textbf{Environment} is extended by two additional wind field components. The components contain scripts to read the measured wind speed and forecasted wind field respectively and spawning trails that move according to local wind speeds and directions. 
                \item The \textbf{GUI} is enhanced with scripts to toggle the visualization of the data on the turbine and the wind trails, and the control panel. Additionally, several scripts are implemented to plot historic, current, and potentially forecasted data of the measured parameters in a conventional 2D graph.
            \end{itemize}

    \subsection{Predictive Digital Twin}
        \label{ssec:i3}
        In this work, predictive capabilities are demonstrated through external weather forecasts, as well as through parameter predictions with data-driven and HAM approaches.
        
        \subsubsection*{External Predictions}
            \label{ssec:i3 external}
            In this demonstration of a predictive DT, weather forecasts from the Norwegian Meteorological Institute\footnote{\url{https://thredds.met.no/}} are being streamed into the DT in real-time through HTTPS requests to their server. In the current implementation, the two-dimensional vector field of wind speed and direction at 10-meter height around the turbine is used, but the framework can easily be extended to also include other forecasted parameters such as temperature, precipitation, relative humidity, and pressure. The streamed data can be visualized in the DT e.g. through trains as shown in Fig.~\ref{fig:trails}.

        \subsubsection*{Neural Networks}
            \label{ssec:i3 internal}
            In addition to the weather forecasts, DNNs and LSTMs were trained on historic data to predict the values from 1-10 seconds, minutes, and hours ahead for seventeen parameters. The predicted parameters include active and reactive power, rotor RPM, wind speed, current and average wave height, blade pitch of each blade, yaw angle, and 6DOF tower motion.

            For the training of the DNNs and LSTMs, the first four months of the data explained in Sec.~\ref{ssec:i1 available} were used. 10\% of this training data was used for validation. The remaining four months were used to test the model. This results in 9.3 million data points for 1-10 seconds forecasts, 155 thousand data points for 1-10 minutes forecasts, and 2.6 thousand data points for 1-10 hours forecasts.
        
\begin{figure}[h]
                \centering
                \includegraphics[width=\linewidth]{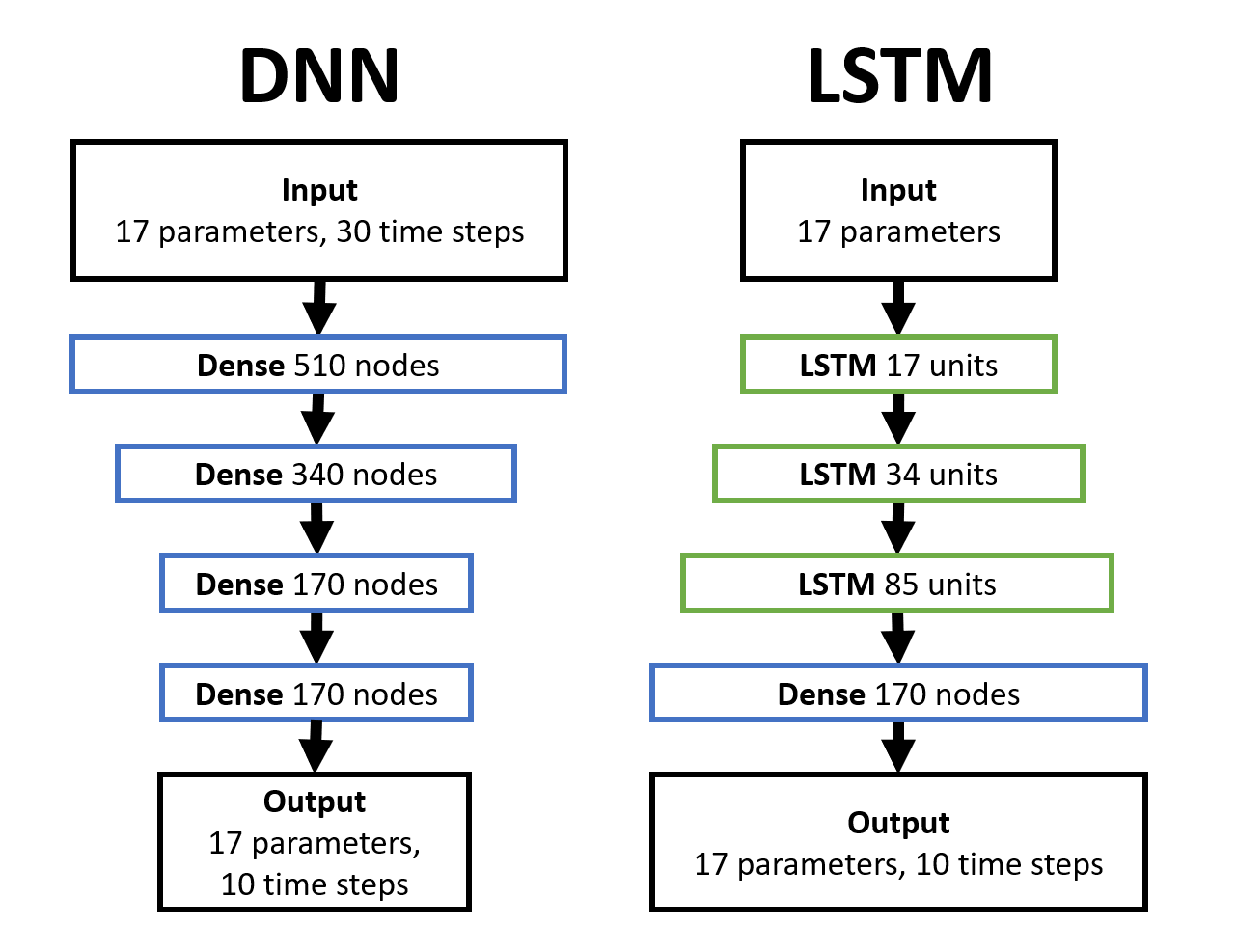}
                \caption{Structures of the DNN and LSTM}
                \label{fig:structure}
            \end{figure}

            Network structures and hyperparameters were tuned manually. The batch size was set to 32, while the last 30 measured values of each parameter were used as input for the predictor. Rectified linear units (ReLU) were chosen as the activation function for both the DNN and LSTM. All parameters are predicted at once. On one hand, strong correlations between environmental-, operational-, and motion-related parameters exist. On the other hand, executing a single NN improves evaluation speed, which is required to perform the predictions in real time on devices with low hardware specs. It is for this reason also that 10 steps are predicted at once.
            Note that the time horizon of the forecast can be further improved by stacking multiple multi-step-ahead forecasts, albeit with reduced accuracy.
            The structures of the DNNs and LSTMs are shown in Fig.~\ref{fig:structure}. The same structure was used across the three timescales.

            The normalized root mean squared errors (NRMSEs) were used to quantify the performance of the model, while the normalized mean squared error (NMSE) was used as the loss function. The MSE of the prediction performed at time $t$ for a value $i$ steps ahead and for one parameter $p$ is therefore
            \begin{equation}
                NMSE_{t,i,p} = \left(\frac{x_{t+i,p} -f_{t+i,p}(x_{t-m+1,p}...x_{t,p})}{\Delta x_p}\right)^2
            \end{equation}
            with $x_{t,p}$ being the value for a specific time $t$ and parameter $p$, $f_{p,t}(x)$ being the prediction for time step $t$ and parameter $p$, $m$ being the number of time steps used as input, and $\Delta x_p$ being the difference between the minimum and maximum value of parameter $p$ in the data set. The NRMSE of one prediction performed at time $t$ can therefore be written as
            \begin{equation}
                NRMSE_t = \sqrt{\frac{\sum_{p=0}^{l}{\sum_{i=0}^{k} {NMSE_{t,i,p}}}}{l k}}
            \end{equation}
            with $k$ representing the number of time steps predicted, and $l$ being the number of parameters in the model.
            Finally, the $NRMSE$ of a prediction model and timescale across the whole data set with length $n$ is simply calculated as
            \begin{equation}
                NRMSE = \sqrt{\frac{\sum_{j=m}^{n-k-1}{NRMSE_{t_j}}^2}{n-m-k}}
            \end{equation}

            The DNN and LSTM for 1-10 seconds forecasts as well as the DNN for 1-10 minutes forecast converged with respect to validation loss within one training epoch. The LSTM for 1-10 minute forecasts took 2 epochs to converge. The validation loss of the DNN and LSTM for 1-10 hour forecasts improved for 3 and 6 epochs respectively before overfitting occurred due to the relatively small amount of individual training samples.

            \subsubsection*{Hybrid Analysis and Modeling}
            
            It became evident that for the hourly forecast, both the DNNs and the LSTMs were performing worse than the persistence model. Comparing the validation and training loss showed that the NNs reached overfitting after only a few epochs. Potential approaches to circumvent this are reducing the size of the NNs, introducing dropout layers, reducing the number of input parameters, gathering more data, or generating mock data. In this work, a HAM approach is used that resembles the last-mentioned method. The NNs are first trained towards emulating a model, in this case, the simple persistence model. The resulting NN weights are then used as the starting point for the training on the actual data, similar to the procedure used in transfer learning. The same approach can be used for more complex physical and parameter-specific models to allow the training of more reliable NNs even with little real data available.
            
            The input samples for this HAM approach were generated with random values between 0 and 1, and the corresponding output was generated by applying the persistence model. The DNNs required 5 million samples to learn the behavior of the persistence model with NRMSE smaller than 0.01, while the LSTM required 10 million samples. 
            This modified training approach allowed the NNs to improve over the persistence model in only a few epochs. Both the pre-trained DNN and LSTM converged after 3 epochs for hourly predictions
            The same HAM approach was also used for minute predictions, where it produced a notable albeit smaller improvement for both network types. A single epoch of training was sufficient to converge with respect to validation loss.

            \subsubsection*{Training Results}
\begin{figure}[h]
                \centering
                \includegraphics[width=\linewidth]{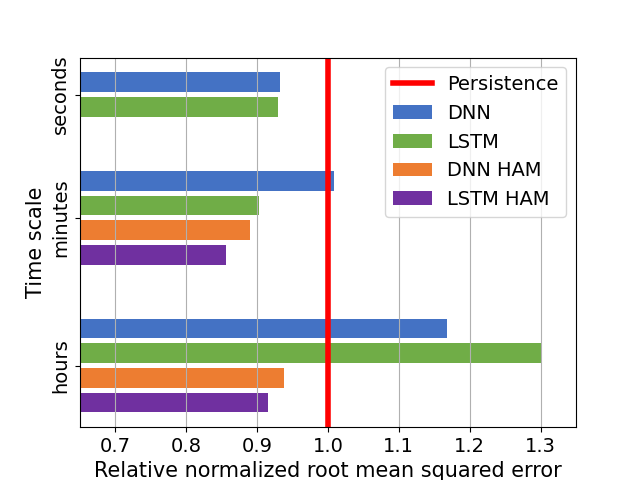}
                \caption{NRMSE of models relative to persistence error. A lower value corresponds to better performance.}
                \label{fig:performance}
            \end{figure}

            The performances of the DNNs and LSTMs, as well as the DNNs and LSTMs with the modified training strategy were benchmarked against the baseline persistence model on the four-month test set. The NRMSE for each model and timescale is shown in Fig.~\ref{fig:performance}.

            For 1-10 second predictions, both NNs are able to leverage the large amount of training data similarly well, even though the LSTM performs slightly better than the DNN. 
            On 1-10 minute forecasts, the HAM approach through pre-training on the persistence model improves the performance of both NNs, with LSTM performing best.
            Both NNs are not able to compete with the persistence model for hourly forecasts. While this may go against intuition at first, note that the hourly forecast offers the least training data. Pre-training the networks on Persistence, however, gave significant improvement for both NNs. Again, the hybrid LSTM performs best of all models.
            
            While it is out of the scope of this work to optimize the NNs beyond this point, further modification of the network structures and combination with more sophisticated models to improve reliability and circumvent overfitting may yield further improvement and will be investigated in future work.
            
        \subsubsection*{Visualization}
        The predictions of wind speed and forecasts streamed from the Norwegian Meteorological Institute are visualized through wind trails as shown in Fig.~\ref{fig:trails}. All parameters predicted with the NNs can be inspected graphically (Fig.~\ref{fig:graphs}) and on the 3D model.

        \subsubsection*{Setup Structure}
            \label{ssec:i3 structure}
            The streaming of the external forecast requires a modification to the \textbf{Data Loader} component to access HTTPS servers, and a modification to the \textbf{Environment} component to visualize the wind trails in a vector field. The NN base predictions happen on the server side for increased performance but can be loaded into the digital twin. Direct implementation on the client side has been tested too but has proven unfeasible so far due to software incompatibility, and is disfavoured due to reduced hardware specifications.
    
\section{Results and Discussions}
In this section, the demonstrated capabilities are summarized, areas of improvement are identified, and future work is outlined. Furthermore, considerations for adaption of the implementation to other offshore assets are given, and an outlook towards the impact of DTs on offshore engineering.
    \label{sec:discussion}
    \subsection{Standalone}
        In the standalone DT, a realistic 3D model of the floating wind turbine was created. The 3D model was imported into a game engine and complemented with an ocean and sky environment. Scripts were added that allow adjusting parameters and visualizing them on the turbine and its environment. In future work, the potential of standalone DTs for other offline applications such as turbine siting will be explored. 
        
    \subsection{Descriptive}
        Diverse sensors on the real floating offshore wind turbine Zefyros measured data ranging from met-ocean conditions over turbine parameters to production information. The data was combined with streamed weather forecasts and the data was visualized in the digital twin in real-time. The result is a data interface that combines and connects all data sources for monitoring and inspecting both real-time and historic data. Without a doubt, there are many additional ways to increase the understandability and correlation of data through visualization, such as heat maps, vector fields, and color-encoded density clouds. Going forward, the monitored parameters will be extended by tower base loads, gearbox vibration, and acoustic measurements inside the turbine nacelle. Those parameters can then be used by a diagnostic component to detect small faults and enable predictive maintenance.
    
    \subsection{Predictive}
        Predictive capabilities have been implemented into the DT by streaming weather forecasts and using DNNs and LSTMs on historic turbine data for short- and mid-term predictions of turbine-specific parameters. Future work will aim at improving the prediction models and including the aforementioned tower loads, which are relevant for both maintenance and control.
        
    \subsection{Other Capability Levels}
        In this work, standalone, descriptive, and predictive capabilities of the DT have been demonstrated. Moving forward, the feasibility of diagnostic and prescriptive capabilities on floating offshore wind turbines will be investigated. However, the availability of data from floating wind turbines is limited both in terms of installed sensors and proprietary conflicts, especially for the estimation of damage, component wear, and RUL. This impacts both the available data for calibrating data-driven methods and the turbine-specific parameters required as input for physics-based models. Therefore, offline wind turbine simulations will have to be used further to develop online models for improved maintenance scheduling and demonstrate the value of alleviating proprietary restrictions. Such studies have already been performed for onshore wind turbines, e.g. in \cite{Branlard2020akf} to estimate the tower loads. Adopting these approaches to floating wind turbines will increase the complexity. HAM approaches combine physics-based and data-driven methods and are expected to result in faster, more accurate, and more reliable models. Through transfer learning, they can be adapted from simulations to real wind turbines.
        
    \subsection{Impact on Offshore Engineering in General}
        This work focuses on floating offshore wind farms. Nonetheless, the concept, features, and implementation can be generalized to other offshore assets and processes and will generate similar values throughout the whole life cycle of the asset.
        
        Specifically, changes to consider for the standalone DT are connected to the CAD model and proprietary rights. For assets with very complex geometry, the CAD model might need to undergo vertex pruning to keep it compatible with real-time rendering. The proprietary rights on geometrical and material models of components can increase the work involved in creating a standalone DT. An increase in collaboration between original equipment manufacturers and asset operators may reduce this work in the future.
        
        Just like the standalone DT, the descriptive DT can be adopted for other offshore assets or processes where relevant data can be measured and can be combined into a multi-asset DT. However, industry-specific standards have to be taken into account when streaming data from multiple sources.
        
        Some of the predicted parameters presented here are specific to wind turbines and the models have been optimized for the Zefyros turbine, but similar data-driven and HAM models can be created to predict relevant parameters for other offshore assets. 
        
        While there are still challenges to address, the potential DTs bring to offshore engineering are manifold across the whole life cycle of any asset or process. Especially O\&M processes can benefit from remote monitoring through condition-based maintenance, predictive maintenance, and virtual inspections. DTs facilitate informed decision-making and allow communication of technical details even with non-technical stakeholders. They provide a collaboration platform for data bundling and sharing across fields, and a data and visualization platform for various analyses. In general, the usage of DTs will enable an increase in the return on investment and the sustainability of offshore assets.

\section{Conclusion}
    \label{sec:conclusion}
    In this work, the definition of digital twins in the context of offshore wind energy was concretized and a general scale to categorize them based on their capability was adopted. The concepts were demonstrated by developing a digital twin of an existing floating offshore wind turbine. A standalone digital twin was developed by creating a 3D CAD model of the turbine in Blender and importing it into the game engine Unity for realistic visualization. The environment consisting of sky and ocean surface at this stage was disconnected from reality. At the next step, the descriptive level, real data from the sensors installed on the turbine was visualized inside the digital twin using an Oculus VR headset in virtual reality. Additionally, meteorological data from weather forecasts was streamed in real-time into the digital twin. At the predictive level, machine learning methods were used to predict parameters into the future. It is important to state that the framework proposed here is highly extendable and can be adapted to accommodate different sources of data and features without jeopardizing the effectiveness of the digital twin at any capability level. 
    

\section*{Acknowledgements}
This publication has been prepared as part of NorthWind (Norwegian Research Centre on Wind Energy) co-financed by the Research Council of Norway (project code 321954), industry and research partners. Read more at \url{www.northwindresearch.no} Data for the project realization was provided by SINTEF Energi AS. Mock data was used for the creation of all figures to comply with confidentiality agreements. Picture of the real Zefyros wind turbine in Fig.~\ref{fig:dt} by Fride Moen, Norwegian Offshore Wind.



\bibliographystyle{preprint}
\bibliography{preprint}

\appendix

\end{document}